\documentclass[aps,prd, twocolumn,floatfix,preprintnumbers,superscriptaddress,showpacs]{revtex4-1}
\usepackage{fontenc}
\usepackage{graphicx}
\usepackage{amsmath,amsfonts,bm}
\usepackage{amssymb}
\usepackage{slashed}
\usepackage{color}

\begin{document}
%


\title{Pion Distribution Amplitude and Quasi-Distributions }

\author{A. V. Radyushkin}

\affiliation{Physics Department, Old Dominion University, Norfolk,
             VA 23529, USA}
\affiliation{Thomas Jefferson National Accelerator Facility,
              Newport News, VA 23606, USA
}

\begin{abstract}

We  extend our analysis of quasi-distributions onto the pion distribution amplitude.
Using  the   formalism of parton virtuality distribution  amplitudes (VDAs),  
we establish a connection between the pion 
transverse momentum dependent 
 distribution   amplitude (TMDA) ${\Psi} (x, k_\perp^2)$ and the pion 
quasi-distribution  amplitude (QDA) $Q_\pi (y,p_3)$.
We build  models for the QDAs from  the VDA-based models for soft TMDAs,  and  analyze 
the $p_3$ dependence of  the resulting QDAs. 
As  there are many models claimed to describe  the primordial 
shape of the pion DA,  we   present  the $p_3$-evolution patterns 
for models producing  some popular proposals: Chernyak-Zhitnitsky, flat and asymptotic DAs.
Our results 
 may be used as a  guide  for future  studies 
 of the pion distribution amplitude   on  the lattice using the quasi-distribution approach.

\end{abstract}

\maketitle



 \section{Introduction}

The parton distribution functions (PDFs) $f(x)$, and two-body distribution amplitudes (DAs) 
$\varphi (x)$ are  related to   matrix elements
of bilocal operators on   the light cone $z^2=0$, which prevents 
a straightforward calculation of these functions in  a  lattice 
 gauge  theory 
 formulated in the Euclidean space. 
 The usual way  out is to calculate  their moments.
 In particular, high precision   lattice  calculations of 
  the second  moment of the pion distribution amplitude
 $\varphi_\pi (x)$  were reported in Ref.   \cite{Braun:2015axa}.   
 However, recently, X. Ji  \cite{Ji:2013dva} suggested a method 
 allowing to calculate PDFs  and DAs as  functions of $x$. 
 To this end, he proposes to use purely space-like
separations $z=(0,0,0,z_3)$.  

The matrix elements of equal-time bilocal operators 
produce   distributions  $Q(y,p_3)$  in the momentum $p_3$  component (quasi-distributions).  
The crucial point is that   they 
   tend to the  light-cone distributions $f(y)$, $\varphi (y)$
     in the \mbox{$p_3 \to \infty$}  limit. 
In case of PDFs,  
the results of lattice calculations of  the  parton 
quasi-distributions (PQDs)  were reported  
in Refs.  
\cite{Lin:2014yra,Alexandrou:2015rja,Lin:2015vxw,Chen:2016utp,Alexandrou:2016bud,Alexandrou:2016tjj}.
It is expected \cite{Xiong:2013bka} that PQDs $Q(y,p_3)$ should  have  
 a mild perturbative evolution 
 \mbox{\cite{Gribov:1972ri,Lipatov:1974qm,Altarelli:1977zs,Dokshitzer:1977sg} } 
with respect to $p_3$  for large $p_3$.  
However,  the values of $p_3$ used in the cited lattice  calculations are not very
large,  and the observed strong  variation of 
PQDs  with $p_3$   
 does not have a perturbative form.

In our recent paper \cite{Radyushkin:2016hsy}
we  have studied  nonperturbative evolution of
PQDs
using the formalism of {\it virtuality distribution functions} 
 \cite{Radyushkin:2014vla,Radyushkin:2015gpa}.
 We found that  PQDs can be obtained from 
  the   transverse momentum dependent 
 distributions    (TMDs) ${\cal F} (x, k_\perp^2)$. Then we  
 built models for the    nonperturbative evolution of
PQDs using simple models for TMDs. 
Our results are in qualitative agreement with the
$p_3$-evolution  patterns  obtained in lattice calculations
\cite{Lin:2014yra,Alexandrou:2015rja,Lin:2015vxw,Chen:2016utp,Alexandrou:2016bud,Alexandrou:2016tjj}
 and also in 
 diquark spectator  models \mbox{\cite{Gamberg:2014zwa,Gamberg:2015opc,Bacchetta:2016zjm}. }

As  emphasized in Ref.  \cite{Radyushkin:2016hsy}, because of the
relation between  PQDs and TMDs, the   nonperturbative evolution of
PQDs reflects the $k_\perp$-dependence of the TMDs 
${\cal F} (x, k_\perp^2)$, and thus its  study provides
a new approach to the investigation 
of the 3-dimensional structure of hadrons. 

Our goal in the present paper is to perform a similar analysis 
of the pion {\it quasi-distribution amplitude} (QDA) $Q_\pi (y,p_3)$ 
that produces the pion DA $\varphi_\pi (y)$ in the \mbox{large-$p_3$}  limit.
The basic ingredients of our analysis are {\it virtual distribution amplitudes} 
and {\it transverse momentum dependent amplitudes} 
introduced in Refs.  \cite{Radyushkin:2014vla,Radyushkin:2015gpa}.
 
The paper is organized as follows.
We start in  Section  2  with   an introductory overview of the basic concepts involved.
First, we remind a covariant definition of the twist-2 pion distribution amplitude. 
After that,  we discuss its definition within   the light-front formalism.  
Then we outline the basics of the VDA/TMDA  approach. In Section 3, 
we discuss the quasi-distribution amplitudes. 
  In particular, we show that QDAs are completely determined
 by TMDAs through a rather simple transformation.
 Since the basic relations between the parton distributions
 are rather insensitive to complications brought by spin,
 in Section  3 we refer to a simple scalar model.
 In Section  4,   we discuss modifications related to  quark spin 
 and gauge nature of gluons in quantum chromodynamics (QCD).
 In Section 5 we discuss the VDA-based models for soft TMDAs,
 and  present our results for  nonperturbative evolution 
 of QDAs obtained in these models. The large-$p_3$ limit of  perturbative evolution 
 is discussed in Section  6. 
 Our conclusions are given in Section 7.

\section{Pion Distribution Amplitude}

\subsection{Covariant Definition}

The pion {\it distribution amplitude}  (DA) $\varphi_\pi (x,\mu^2)$ was originally   introduced 
\cite{Radyushkin:1977gp}  as a function
$ \varphi_\pi  (x,\mu^2)$   whose $x^n$ moments 
\begin{align}
f_n (\mu^2) =  \int_0^1 x^n \, \varphi_\pi  (x,\mu^2) \,  dx \, 
\label{covdef}
\end{align}
are  given by reduced matrix elements of  twist-2 local  operators 
\begin{align}
i^{n+1}  R_{\mu^2} &
 \left \langle 0 | \bar d (0) \gamma_5 \left \{  \gamma_\nu D_{\nu_1} \ldots D_{\nu_n} \right \}  u (0) | \pi^+, P 
\right \rangle \nonumber \\ &
=  \left \{ P_\nu P_{\nu_1} \ldots P_{\nu_n} \right \} \, f_n (\mu^2)  \  .
\end{align}
As usual  $\{ \ldots \}$ denotes 
the  twist-2 projection of a Lorentz structure, i.e., symmetrization of indices and subtraction of traces.
Since matrix elements of local operators with $n >0$ diverge, 
one needs to supply them by a renormalization procedure 
denoted above by $R_{\mu^2} $, with $\mu^2$ being the renormalization scale.
In QCD, the   standard choice  of    $R_{\mu^2} $ is based on the dimensional regularization 
and the modified minimal subtraction scheme  $\overline{\rm MS}$.
As a result of such a renormalization,    the zeroth   moment
$f_0(\mu^2)$  does not have 
 $\mu^2$-dependence 
 since the anomalous dimension
of  the axial current is zero. Hence,  $f_0(\mu^2)$ 
for all $\mu^2$  is  equal to  the pion decay constant $f_\pi$
\begin{align}
 f_0(\mu^2) =  \int_0^1  \varphi_\pi  (x,\mu^2) \,  dx = f_\pi \ 
\label{fpi}
\end{align}
 known experimentally,  $f_\pi \approx 130$\,MeV. 

This  definition of DA is  oriented on  
the  use of  the operator product expansion and 
a  description of  the pion  in terms  of the twist-2 DA $\varphi_\pi (x, \mu^2)$ 
that  gives  the collinear distribution of the pion momentum $p$
 among its two valence constituents. The dependence of 
 $\varphi_\pi (x, \mu^2)$ on $\mu^2$  is 
 governed by  perturbative  evolution 
 \cite{Efremov:1978rn,Efremov:1979qk,Lepage:1979zb}
 and does not reflect the primordial (nonperturbative)
 pion's structure in  the  direction  transverse to $p$.
 
 As is well-known, for very large $\mu^2$, the pion DA 
 tends to the ``asymptotic DA''  \mbox{$\varphi^{\rm as}_\pi  (x) = 6 f_\pi x (1-x)$}  
 \cite{Efremov:1978fi}. 
 In general,  $\varphi_\pi (x, \mu^2)$ may differ from its asymptotic form.
 Over the years, several forms were proposed for the pion DA ``at low normalization point'', e.g.,  
Chernyak-Zhitnitsky DA  \mbox{$\varphi^{\rm CZ}_\pi   (x) = 30 f_\pi x (1-x)(1- 2x)^2$}     \cite{Chernyak:1981zz},
 ``flat DA''   $\varphi^{\rm flat}_\pi   (x) =f_\pi$ \cite{Dittes:1981aw,Anikin:1999cx,RuizArriola:2002bp,Radyushkin:2009zg,Polyakov:2009je}, ``root DA'' $\varphi^{\rm root}_\pi   (x) =8 f_\pi \sqrt{x (1-x)}/\pi$
 \cite{Mikhailov:1986be}, etc.

\subsection{Light-Front Formalism  Definition}

A  different  definition \cite{Lepage:1979zb} is used 
in the light-front (LF) quantization framework, where the 
pion distribution amplitude $ \phi_\pi  (x,\mu^2)$  is understood as  the $k_\perp$-integral   
\begin{align}
  \phi_\pi  (x,\mu^2) = \frac{\sqrt{6}}{(2 \pi)^3} \, 
\int_{k_{\perp}^2\leq \mu^2}  \psi(x,k_{\perp}) \, d^2 k_{\perp}
\label{phiLB}
\end{align}
of the  light-front   wave function (LFWF)  $\psi(x,k_{\perp})$. 
We intentionally use here  a different notation $  \phi_\pi  (x,\mu^2)$
 to emphasize the  fact that      $\psi(x,k_{\perp})$ is 
an object of the  {\it Hamiltonian light-front framework}, while 
 the pion DA   $  \varphi_\pi  (x,\mu^2)$ in Eq. (\ref{covdef}) is defined 
 within the {\it covariant  Lagrangian 
formulation}  of the   quantum field theory (QFT). 

Another difference is the use of a  straightforward cut-off $k_\perp^2 \leq \mu^2$
rather  than a more sophisticated $\overline{\rm MS}$-like subtraction.
As a result, $ \phi_\pi  (x,\mu^2)$ has a nonperturbative evolution with $\mu^2$ even if the 
perturbative evolution is absent. 
Take  a simple example  $\psi (x, k_\perp)\sim \phi (x)  \, e^{-k_\perp^2/\Lambda^2}$. 
Then the zeroth $x$-moment  of 
$\phi_\pi  (x,\mu^2)  $ has the \mbox{$\sim [1- e^{-\mu^2/\Lambda^2}]$}-dependence,
i.e. it is not constant,  reaching  $f_\pi$ in the $\mu^2  \to \infty$ limit only.

Of course, if one has in mind only the applications 
in which  nonperturbative part of the $\mu^2$-dependence may be ignored,
then $\phi_\pi  (x,\mu^2)  $ of the LF definition is very similar 
to the covariantly  defined $\varphi_\pi  (x,\mu^2)  $, and 
the difference between them may be treated as the  use 
of  different renormalization schemes. 

As a matter of fact,  in actual LF calculations 
one encounters LFWFs integrated to some process-dependent scale $\mu$,
i.e. the choice of the renormalization prescription and the scale $\mu$ 
is dictated by diagrams.
Moreover, if the relevant $\mu^2$'s are not extremely large,
the  simple example above shows that one may need to take into account
the nonperturbative $\mu^2$-dependence 
of   $\phi_\pi  (x,\mu^2)  $   reflecting  the transverse momentum 
behavior of  the LFWF $\psi (x, k_\perp)$, i.e., the 3-dimensional structure
of the pion, which may be essential for some  processes.

In particular, the photon-pion transition form factor involves 
 \mbox{$\phi_\pi (x, \mu^2=x^2 Q^2)/[xQ^2]$,} i.e.  
 LFWF $\psi (x, k_\perp)$ integrated over $k_\perp $ till $xQ$
 \cite{Lepage:1980fj,Musatov:1997pu}. 
As a result,  the remaining $x$-integral 
in the  LF formula has a finite $Q^2 \to 0$ limit:  the 
infrared small-$x$ divergence is eliminated by 
a cut-off provided by $\phi (x, \mu^2=x^2 Q^2)$.
On the other hand,  a  formula involving $\overline{\rm MS}$-based DA
$\varphi (x, \mu^2)$ with a fixed scale $\mu^2$ is singular in 
the $Q^2 \to 0$ limit. 
One may question  the applicability of the LF formula 
down to  $Q^2=0$, but at least it does not give 
an infinite result  for a quantity that is known to be finite. 
For  this reason, the  LF formula  looks as a more attractive  tool for
modeling the form factor behavior at 
moderate $Q^2$ than the perturbative QCD $1/Q^2$   twist expansion.

Still, a problem with the LF formalism is  that 
 LFWFs     are  not directly connected
with the usual objects of the covariant field theory, such as matrix elements
of local or nonlocal operators.

 In our papers  \cite{Radyushkin:2014vla,Radyushkin:2015gpa}, we have developed 
 the formalism of
 {\it virtuality distribution amplitudes} (VDAs) 
 that is  fully based on the covariant field theory concepts. In the VDA approach, 
 the pion  is described 
 by the 
  {\it transverse momentum dependent distribution
amplitude} (TMDA)  which 
 has a  direct  connection  with the objects of the covariant QFT.
On the other hand, just like the LF wave functions, the 
\mbox{TMDAs}   give a 3-dimensional  description of the pion structure.  
 
\subsection{Pion TMDA}

To omit  inessential    complications related to spin,
 we  illustrate the ideas underlying  TMDAs using  
  a simple example of a  scalar  theory.
The  key element of our  approach 
 \cite{Radyushkin:2014vla} is the  {\it VDA representation}   
 \begin{align}
 \langle 0 |   \psi(0) \psi (z)|p \rangle 
= & 
\int_{0}^{\infty} d \sigma \int_{0}^1 dx\,   \nonumber \\ & \times 
 \Phi (x,\sigma) \,  \,  e^{-i x (pz) -i \sigma (z^2-i \epsilon)/4 } \,   
 \label{Phixs000}
\end{align} 
that basically reflects the fact that the matrix element 
$\langle 0 |   \psi(0) \psi (z)|p \rangle$  depends on $z$ through 
$(pz)$ and $z^2$. It may be treated as a double Fourier representation 
with respect to these variables.

The main 
     non-trivial   feature of this representation  
is  in its  specific  limits of integration over  $x$ and  $\sigma$.  
They hold  for any contributing Feynman diagram  \cite{Radyushkin:2015gpa}, 
so we assume that this property
is true in general. 
Note that starting with the first loop, the diagram contributions are non-analytic in $z^2$
due to $\ln z^2$ factors,
but   the VDA representation,  unlike the Taylor expansion  in $z^2$,   is valid nevertheless.

While the VDA  representation  is  a fully covariant expression, it is convenient 
to use a frame in which  the  pion momentum $p$ is  purely longitudinal 
$p= (E, {\bf 0}_\perp, P)$. 
Choosing some special cases of $z$, one can get representations
for several parton functions, all in terms of one and the same universal VDA $ \Phi (x,\sigma) $.
In particular, choosing  $z$  on the light front $z_+=0$ and with $z_\perp=0$
(i.e., taking $z=z_-$)
 gives the twist-2 distribution amplitude $\varphi (x)$
  \begin{align}
  \langle  0 |\psi (0) \psi(z_-)  
| p \rangle =
   \int_{0}^1 dx \, \varphi  (x) \, 
e^{-ixp_+ z_-} \,  \   . 
 \label{twist2par0}
\end{align}
Comparing this relation with the VDA representation 
we have 
\begin{align} 
 \varphi  (x)   = \int_{0}^{\infty}  \Phi (x,\sigma) \, d \sigma  \ , 
\label{Phix0}
\end{align} 
provided that 
the 
  $z^2 \to 0$ limit is finite, e.g. in the  super-renormalizable
  $\varphi^3$ theory.  In   the  renormalizable
  $\varphi^4$  theory, the function $ \Phi (x,\sigma)$ has a $\sim 1/\sigma$ hard 
  part, and the integral (\ref{Phix0}) is logarithmically divergent,
  reflecting the perturbative evolution of the DA  in such a theory.
  In this case, one may arrange a 
   regularization of the $\sigma$-integral characterized   by some parameter $\mu^2$.
   Then $\varphi (x) \to \varphi (x,\mu^2)$.

Light-cone singularities  are avoided   if we   choose a spacelike $z$, e.g.,  take 
 $z$ that has 
 $z_-$ and $z_\perp$ components only. Then we   can  
   introduce the  {\it transverse momentum dependent distribution  amplitude}   
   $  {\Psi}(x, k_\perp^2 )$  
   as a Fourier transform 
   \begin{align}
  \langle  0 |\psi (0) &\psi(z_-,z_\perp)  
| p \rangle =  \int_{0}^1 dx \, 
 \, e^{-i x p_+ z_- }   \nonumber \\ & \times  \int  {d^2 k_\perp }   {\Psi}(x, k_\perp^2 ) \, e^{i (k_\perp z_\perp)}
\, \  
\,  \  
 \label{tmda}
\end{align}
of the matrix element with respect to
   $z_-$  and $z_\perp$. 
Because of   the rotational 
invariance in $z_\perp$ plane,   TMDA  depends on $k_\perp^2$  only, the 
fact  already  reflected in  the notation. 
   The TMDA may be written in terms of the  VDA as  
\begin{align}
{\Psi} (x, k_\perp^2 ) = & \frac{i }{\pi }
\int_{0}^{\infty} \frac{d \sigma }{\sigma} \, 
 \Phi (x,\sigma) \,  \,  
 e^{- i (k_\perp ^2-i \epsilon )/ \sigma} 
  \  .  \label{TMDsig} 
\end{align} 
The integrated TMDA   
 \begin{align}
{\mathfrak f }  (x, \mu^2) \equiv \pi   \int^{\mu^2}_0 d k_\perp^2   \Psi (x, k_\perp^2) 
\label{intTMD}
\end{align}  
is analogous to the $\mu^2$-dependent pion distribution amplitude
$\phi (x, \mu^2)$
of the  LF  formalism (but,  of course, 
being an object of  the covariant QFT, ${\mathfrak f}(x, \mu^2)$   does not coincide with it). 
In terms of the VDA, 
\begin{align}
&{\mathfrak f} (x, \mu^2)   = 
\int_{0}^{\infty} {d \sigma }\, \left [ 1 - e^{- i (\mu^2-i \epsilon )/\sigma} \right ]\, 
 \Phi (x,\sigma) \,  \  .  
\label{intTMDsig} 
\end{align} 
Since it is defined by a straightforward cut-off,  ${\mathfrak f}(x, \mu^2)$ 
{\it evolves}
with $\mu^2$ even if the limit $\mu^2 \to \infty$ is finite, e.g. in a super-renormalizable 
theory.  The evolution equation 
\begin{align}
\mu^2 \frac{d}{d \mu^2}  {\mathfrak f}  (x, \mu^2)   = & \pi \mu^2 \Psi (x, \mu^2) \  
\label{intTMDevo} 
\end{align} 
follows from the definition (\ref{intTMD}). 
When the TMDA $\Psi  (x, k_\perp^2)$  vanishes faster than 
$1/k_\perp^2$ (such a TMDA will be called ``soft''),  evolution 
essentially stops at large $\mu^2$. 

In a renormalizable theory, it makes sense to treat 
$ \Phi (x,\sigma)$  as a sum of a soft part $ \Phi^{\rm soft}  (x,\sigma)$, 
generating a nonperturbative evolution of ${\mathfrak f}(x,\mu^2)$, 
and a $\sim1/\sigma$ hard tail.  
To avoid nonperturbative evolution, one may choose 
an $\overline{\rm MS}$-type construction, e.g. regularize the $\sigma$-integral
in Eq. (\ref{Phix0})  
by a $\sigma^{-\epsilon}$ factor and then subtract $1/\epsilon$ poles.

However, just like in the LF formalism, the objects that appear
 in  actual
calculations are exactly the integrated TMDAs 
rather than their  $\overline{\rm MS}$-type sisters. 
 In particular, the photon-pion transition form factor
is given in the VDA approach by the $x$-integral of
${\mathfrak f} (x, \mu^2)/[xQ^2]$  taken at $\mu^2 = x Q^2$
\cite{Radyushkin:2014vla}, i.e., it  involves 
TMDA  $\Psi (x, k_\perp^2)$ integrated over $k_\perp^2 $ till $xQ^2$.
As a result,  the TMDA  formula has a finite $Q^2 \to 0$ limit. 
Furthermore, using simple models for soft  TMDAs  one can get a very 
close description of experimental data by the nonperturbative evolution 
of the integrated   TMDA \cite{Radyushkin:2015gpa}.   

For very large $\mu^2$, the perturbative evolution dominates   and 
eventually brings ${\mathfrak f} (x, \mu^2)$  to its {asymptotic form}
 $6 f_\pi \, x (1-x)$.
The question, however, is what  kind of shape   ${\mathfrak f} (x, \mu^2)$ has 
at low  scales $\mu \sim 1$ \,GeV, and also how this shape changes with $\mu^2$. 
As we have discussed, this nonperturbative $\mu^2$-evolution 
reflects the $k_\perp$  dependence  of the soft part of the pion TMDA.

Below, we shall see that there is another function, the pion 
{\it quasi-distribution amplitude}  $Q_\pi (y, P)$ whose \mbox{$P$-dependence } 
is  also determined by the $k_\perp$-dependence   of the pion TMDA. 
The quasi-distributions  have been introduced recently by X. Ji  \cite{Ji:2013dva} 
to facilitate a calculation of light-front functions (PDFs, DAs, etc.) on the lattice.

\section{Quasi-Distribution Amplitude} 

\subsection{Definition} 

The basic proposal  of Ref. \cite{Ji:2013dva} is to consider equal-time bilocal operators 
corresponding to 
$z= (0,0,0,z_3)$ [or, for brevity, \mbox{$z=z_3$}].  Incorporating the VDA representation,
we have  
 \begin{align}
  \langle 0 |   \psi(0) \psi (z_3)|p \rangle 
=  & 
\int_{0}^{\infty} d \sigma \int_{-1}^1 dx\,  %
 \Phi (x,\sigma) \, 
 e^{i x p_3 z_3 +i \sigma z_3^2/{4}} \,  .
 \
 \label{newVDFxz3}
\end{align} 
Using  again the   frame in which $p=(E, 0_\perp, P)$, 
 and 
introducing  the pion quasi-distribution  amplitude   through 
 \begin{align}
  \langle 0 |   \psi(0) \psi (z_3)|p \rangle 
=  & 
\int_{-\infty}^{\infty}   dy \, 
 Q_\pi (y, P) \,  e^{-i y  P z_3 } \,  , 
 \
 \label{newVDFxzQ}
\end{align} 
we get a relation between QDA and VDA,
 \begin{align}
 Q_\pi (y, P)  = &  \sqrt{\frac{i \, P^2}{\pi }} \,\int_{0}^{\infty} 
 \frac{ d   \sigma}{\sqrt{ \sigma}}
 \int_{0}^1 dx\,  %
 \Phi (x,\sigma) \, 
  e^{- i (x -y)^2 P^2 / \sigma }
 \ . 
 \label{newVDFzQin2}
\end{align} 
It is easy to see that, for large $P$,
we have 
 \begin{align}
  \sqrt{\frac{i\, P^2}{\pi \sigma}}
  e^{- i (x -y)^2 P^2 / \sigma } =  \delta (x-y) + \frac{\sigma}{4 P^2} \delta'' (x-y) + \ldots
 \
 \label{Qin3}
\end{align} 
and $Q_\pi (y, P\to \infty)$ tends to the integral  (\ref{Phix0})  leading to  $\varphi_\pi (y)$. This observation 
suggests that one may be able to  extract the ``light-cone''  distribution 
amplitude $\varphi_\pi  (y)$
from the  studies of  the purely ``space-like''   function  $Q_\pi (y, P)$ for large $P$
 \cite{Ji:2013dva}.

\subsection{Evolution}

Again, to  study the $P$-evolution of $Q_\pi (y, P)$
it makes sense to  split  $  \Phi (x,\sigma) $  into the soft part, for which the integral over $\sigma$ 
is finite,
and the hard tail that generates perturbative evolution. 

The nonperturbative evolution of $Q^{\rm soft} (y,P)$ with respect to $P$ has the area-preserving property.
Namely, since 
 \begin{align}
 \int_{-\infty}^\infty dy\, 
  e^{- i(x -y)^2 P^2 / \sigma } %
  = \sqrt{\frac{\pi \sigma}{i P^2}}
 \
 \label{Qin71}
\end{align} 
we  have 
 \begin{align}
 \int_{-\infty}^\infty dy\,  Q_\pi^{\rm soft}(y, P)  = &
  \int_{0}^1 dx \, \varphi_\pi^{\rm soft} (x) =f_\pi 
   \  .
 \
 \label{Qin810}
\end{align} 
In other words, 
 $Q_\pi^{\rm soft} (y,P)$  for any $P$ 
has  the same area normalization as $\varphi _\pi^{\rm soft}(x)$.  
In this respect, the pion QDA  pleasantly differs from the 
integrated TMDA ${\mathfrak f}^{\rm soft} (x, \mu^2)$  whose zeroth moment
is $\mu^2$-dependent.

Similarly, we have the momentum sum rule
 \begin{align}
 \int_{-\infty}^\infty dy\,  y\, Q_\pi^{\rm soft} (y, P)  = 
  \int_{0}^1 dx\,  x \, 
\varphi_\pi^{\rm soft} (x)  \  .
 \
 \label{Qin812}
\end{align} 

\subsection{Relation to TMDA} 

Comparing the VDA representation (\ref{Qin3})  for $Q_\pi(y, P)$ with that for the TMDA
$\Psi (x, k_\perp^2)$  (\ref{TMDsig}) 
(note that they  are valid  both  for soft and hard parts)  we conclude that
  \begin{align}
 Q_\pi (y, P)  =  & \,\int_{-\infty}^{\infty} d  k_1
   \int_{0}^1 dx\, P\,  \Psi (x, k_1^2+(x -y)^2 P^2 )
  \  .
 \label{QTMD}
\end{align} 
Thus, the quasi-distribution amplitude 
 $ Q_\pi (y, P)$ 
is  completely determined by the form
of the TMDA  $\Psi (x, k_\perp^2)$. 

This   formula may be also obtained if one takes
\mbox{$z=(0,z_1,0,z_3)$}  in the VDA representation and 
introduces the momentum $k_1$ conjugate to  $z_1$.
Then 
   \begin{align}
& \int_{-\infty}^{\infty}   dy \, 
 e^{i y  P z_3 }   \langle  0 |\psi (0)  \psi(z_1, z_3)  
| p \rangle  \nonumber \\ &  =  \int_{-\infty}^{\infty} {d k_1 } 
 \, e^{-i k_1 z_1 }    \int_{0}^1 dx \,   {\Psi}(x, k_1^2+(x-y)^2 P^2 )  \  .
 \label{qpd13}
\end{align}
Taking $z_1=0$ gives Eq. (\ref{QTMD}).
Furthermore, introducing  the variable $k_3 \equiv (x-y) P$, we have 
  \begin{align}
 Q_\pi (y, P)  =  & \,\int_{-\infty}^{\infty} d  k_1
   \int_{-yP}^{(1-y)P}  dk_3\,  \Psi (y+k_3/P, k_1^2+k_3^2 )
  \  .
 \label{QTMD2}
\end{align} 

Thus, $Q_\pi (y, P)$ is given by  an integral   over a stripe of width $P$  in the  2-dimensional 
$(k_1,k_3)$  plane.    
When   $P\to \infty $  for a  fixed  nonzero $y$,  the stripe covers the whole 
$(k_1,k_3)$  plane.   Moreover, for a soft TMDA $\Psi (x, k^2)$ that  rapidly decreases outside 
a region $k^2 \lesssim \Lambda^2$, only the values of $k_3 \lesssim \Lambda $ are  
essential, and for large $P$ one may approximate  the first argument of the TMDA by $y$.
Hence, the \mbox{$P \to \infty$}  limit gives $\varphi_\pi ^{\rm soft} (y)$.

For comparison, the integrated TMDA  ${\mathfrak f} (y, \mu^2)$  is obtained
by integrating $\Psi (y, k_\perp^2)$  over a  circle of  radius $\mu$
in the $k_\perp$  plane.  Again, the circle covers the whole plane when $\mu \to \infty$,
and  ${\mathfrak f} ^{\rm soft}  (y, \mu^2) \to \varphi_\pi^{\rm soft} (y)$. 

Thus, while the patterns of the nonperturbative evolution of 
$Q_\pi^{\rm soft} (y, P)$  and ${\mathfrak f} ^{\rm soft}  (y, \mu^2)$  are
different,
they  become more  and more close for large $P$ and $\mu$,
eventually producing the same function $\varphi_\pi^{\rm soft} (y)$.

\section{QCD}

\subsection{Spinor quarks}

In spinor case, one deals with the matrix element   
    \begin{align}
 B^\alpha  (z,p) \equiv \langle  0 | \bar \psi (0) \gamma_5
 \gamma^\alpha  \psi(z)  | p \rangle \  \ .
\end{align}
It may be decomposed into $p^\alpha$ and $z^\alpha$ parts:
$B^\alpha  (z,p) = p^\alpha B_p (z,p) + z^\alpha B_z (z,p)$, or in the VDA 
representation 
\begin{align}
& B^\alpha  (z,p)
  = 
\int_{0}^{\infty} d \sigma \int_{-1}^1 dx\,  \nonumber \\ & \times 
\bigl [  p^\alpha \Phi (x,\sigma) 
+z^\alpha Z (x,\sigma) \bigr ] \, 
 \,  e^{-i  x (pz) -i \sigma {(z^2-i \epsilon )}/{4}} \ . 
 \label{OPhixspin0}
\end{align} 

If we take  $z=(z_-, z_\perp)$ in the $\alpha=+$  component of 
${\cal O}^\alpha$,  the purely higher-twist $z^\alpha$-part drops out 
and we can introduce the TMDA  $\Psi (x, k_\perp^2)$  that 
is  related to the VDA $\Phi (x,\sigma)$ by the scalar formula 
(\ref{TMDsig}). 

In the QDA case, the easiest way to avoid the effects of the $z^\alpha$ admixture   
 is to take the time component of \mbox{$B^\alpha  (z=z_3,p)$}  and define
 \begin{align}
& B^0   (z_3,p)   
  =  p^0 
 \int_{-1}^1 dx\,  
Q_\pi (y,P) \, 
 \,  e^{i  y Pz_3 }  \  . 
 \label{OPhixspin12}
\end{align} 
 
The connection between $Q_\pi (y,P)$  and $\Phi (x, \sigma)$  is  given then by the 
same formula (\ref{newVDFzQin2}) as in the scalar case.  As a result,
we   have the sum rules  (\ref{Qin810}) and (\ref{Qin812})  corresponding 
to charge and momentum conservation. 
Furthermore, the quasi-distribution amplitude  $Q_\pi (y,P)$ is  related
to TMDA  $\Psi (x, k_\perp^2)$ by the scalar conversion formula (\ref{QTMD}).

\subsection{Gauge fields}

In QCD, for $\pi^+$ one should take the operator 
\begin{align}
{\cal O}^\alpha  (0,z; A) \equiv \bar d (0) \,\gamma_5 \, 
 \gamma^\alpha \,  { \hat E} (0,z; A) u (z)  \  
\end{align}
involving a  straight-line 
 path-ordered exponential
\begin{align}
{ \hat E}(0,z; A) \equiv P \exp{ \left [ ig \,  z_\nu\, \int_0^1dt \,   A^\nu (t z) 
 \right ] }  
 \label{straightE}
\end{align}
in the quark (adjoint) representation. 
As is well-known, its   Taylor expansion   has 
the same structure 
as that for the original $\bar \psi (0) \gamma_5  \gamma^\alpha \psi (z)$ 
operator, with the only change   that
one should use covariant derivatives 
\mbox{$D^\nu =\partial^\nu  - ig A^\nu$}  instead of the
ordinary  $\partial^\nu $ ones.

Again, the  $z^\alpha$   admixture  is avoided if  the pion quasi-distribution amplitude 
is    defined through the time component of ${\cal O}^\alpha$. 
Then we 
have the same relation  between the VDA  and QDA  as in the scalar case.
Due to Eq. (\ref{Qin810}), this  
 results   in the area preserving property for the soft part 
 \begin{align}
& \int_{-\infty}^\infty dy\,  Q^{\rm soft} (y, P) =f_\pi 
 \ .
 \label{Qin815}
\end{align} 
Also, due to Eq. (\ref{Qin812})    we  have the momentum sum rule
 \begin{align}
 & \int_{-\infty}^\infty dy\,  (y-1/2) \, Q^{\rm soft} _q(y, P)  =0 \ . 
 \
 \label{Qin8123}
\end{align} 

Since the VDA $\Phi (x,\sigma)$ 
is defined through the matrix element of a  gauge-invariant
operator, it is gauge-invariant also. 
For this reason, TMDA $\Psi (x, k_\perp^2)$  is   a gauge-invariant object
as well.
It  should not be confused with the $k_T$-dependent
(and gauge-dependent)  
``underintegrated  distributions''  that appear 
in perturbative  loop calculations based on Sudakov decomposition 
of  the  integration momentum $k$.

\section{Models for soft part} 

\subsection{Models} 

To get an idea about patterns of the nonperturbative
evolution of the QDAs, we need   some explicit  models  of the 
$k_\perp$ dependence of  soft TMDAs    ${\Psi} (x, k_\perp^2)$.
We will  use here the same models as in our papers 
\cite{Radyushkin:2015gpa,Radyushkin:2016hsy}. 
While TMDAs  are   functions
of two independent variables $x$ and $k_\perp^2$,  we  take, 
for simplicity,  the  case of     factorized models
  \begin{align}
{\Psi} (x, k_\perp^2) = \varphi_\pi (x) \, \psi (k_\perp^2)  \  ,
\label{Ans}
\end{align} 
in which   $x$-dependence and  $k_\perp$-dependence appear in  separate factors.

If we  assume  a Gaussian dependence on
$k_\perp$, 
  \begin{align}
{\Psi} ^{\rm G} (x, k_\perp^2) = \frac{\varphi_\pi (x) }{\pi \Lambda^2}  e^{-k_\perp^2/\Lambda^2} \ , 
\label{Gaussian}
\end{align}  
the  conversion formula (\ref{QTMD}) 
results in 
 \begin{align}
 Q_\pi^{\rm G} (y, P)  = &\frac{P}{\Lambda \sqrt{\pi} }  \,
 \int_{0}^1 dx\,  %
\varphi_\pi (x)  \, 
  e^{- (x -y)^2 P^2 / \Lambda^2 }
 \  . 
 \label{QinG}
\end{align}

 In the space of  impact parameters $z_\perp$, the   Gaussian 
 model gives a $e^{-z_\perp^2\Lambda^2/4}$ fall-off 
 that 
is too fast  for large $z_\perp$. 
As an alternative extreme case, we  take a model with 
the $1/(1+z_\perp^2 \Lambda^2/4)$ dependence on $z_\perp$, whose  
fall-off at large $z_\perp$ is too { slow}.   
 It  corresponds to the ``slow'' model for the TMDA 
\begin{align}
& {\Psi}^{\rm S}    (x, k_\perp^2) = 2  \varphi_\pi (x)  \,  \frac{K_0 ( 2 |k_\perp| / \Lambda) }{ \pi  \Lambda^2 } 
\  
\end{align}
that has a logarithmic singularity for small $k_\perp$
reflecting  a too slow  fall-off  
for large  $z_\perp$.   
For the QDA,  we have 
 \begin{align}
 Q_{\pi} ^{\rm S}     (y, P)  = &\frac{P}{\Lambda}  \,
 \int_{0}^1 dx\,  %
 \varphi_\pi (x)  \, 
  e^{-2 |x -y| P / \Lambda } \  .
 \
 \label{Qinm}
\end{align} 

Note that   the Gaussian model and the ``slow''  model   have the same 
$\sim (1-z_\perp^2 \Lambda^2/4)$  behavior for small $z_\perp$, 
i.e. they correspond 
to the same value of the $\langle 0 | \varphi (0) \partial^2 \varphi (0)| p \rangle$
matrix element (in the scalar case), 
provided that one takes the same value of $\Lambda$ in both models.
  For large $z_\perp$, however,  the Gaussian model  has a  fall-off  that  is 
   too fast, while  the fall-off of the ``slow'' model  is too slow. Thus, they  look like 
 two extreme  cases, and 
provide a good   illustration of the nonperturbative evolution 
of the pion QDA, with  expectation  that other models 
would  produce results somewhere in  between these two  cases.

\subsection{Numerical Results}

To compare evolution patterns induced by the Gaussian and ``slow'' models,
we  take the  Ansatz  (\ref{Ans})  with $\varphi_\pi (x)$  having 
a drastic shape of the Chernyak-Zhitnitsky DA 
\mbox{$\varphi^{\rm CZ}_\pi  (x) = 30 f_\pi x (1-x)(1- 2x)^2$}. 
As one can see from Fig.~\ref{CZ}, for $P/\Lambda =1$ the Gaussian model 
shows no indication of humps visible for higher $P/\Lambda$ ratios.
In the ``slow'' model, small humps are present even for $P/\Lambda =1$.
For high ratios $P/\Lambda =5$ and 10, the two models give close results,
with strong humps. 

Assuming  $\Lambda \sim 0.6$ GeV suggested by the VDA-based  fits
of the photon-pion transition form factor in Ref.  \cite{Radyushkin:2015gpa},
we expect that $P\sim 3$ GeV would be required  to support (or rule out)
the  CZ-type shape of the pion DA.

It is also interesting  to note that the nonperturbative evolution pattern
here is exactly opposite to the perturbative one. In the latter case,
the humps of the initially CZ-shaped DA become less pronounced 
as the normalization scale   increases 
and eventually disappear, with  the  DA  tending  to the asymptotic 
$\sim x (1-x)$ shape. 

To compare patterns of the QDA's  nonperturbative  evolution  for different shapes of the limiting DA,
we take three  models for $\varphi_\pi (x)$: Chernyak-Zhitnitsky $\varphi^{\rm CZ}_\pi  (x)$,
flat \mbox{$\varphi^{\rm flat}_\pi  (x) =  f_\pi $} and asymptotic 
\mbox{$\varphi^{\rm as}_\pi  (x) = 6 f_\pi x (1-x)$}. 
The results in the Gaussian and the ``slow'' models are rather   similar.
 To avoid plotting too many graphs, we take, for definiteness, the ``slow'' model.
 Then,  for the  flat  limiting DA  we have  
 \begin{align}
 \frac1{f_\pi} &
 Q_{\pi}^{\rm S, flat}   \left   (y, P \right )  =  \frac{P}{\Lambda}  \,
 \int_{0}^1 dx\,  %
  e^{-2 |x -y| P / \Lambda }  \  .
 \
 \label{Qinf}
\end{align} 

This integral can  be calculated  analytically.  
Writing $y=(1+\eta)/2 $ in terms of  a symmetric variable  $\eta$, we get 
 \begin{align}
 \frac1{f_\pi} 
 Q_{\pi}^{\rm S, flat}   \left   (y, P \right )  &
  = \left (1- e^{-P/\Lambda} \cosh (P\eta /\Lambda)  \right ) \theta (|\eta|\leq 1) 
    \nonumber \\ & 
    +  \sinh (P/\Lambda) e^{-P|\eta|/\Lambda}  \theta (|\eta|\geq 1) 
  \  .
 \
 \label{Qinmout}
\end{align} 
Similar, but more lengthy expressions may be obtained for two other models. 
As one can see from Fig.~\ref{Qfy13},  for small $P=\Lambda$  we have 
very close  curves.    For  larger $P=3\Lambda$ the difference 
becomes visible, and for large $P=5\Lambda$ and $P=10\Lambda$ 
the curves shown in Fig.~\ref{Qfy510} are   distinctly different.
In  fact, the $P=10\Lambda$ curves are very close to their limiting forms.
Again, the nonperturbative evolution pattern in  case of the flat DA 
is opposite to the perturbative one: as $P$ increases, $Q_\pi^{\rm flat} (y,P)$ broadens  from a rather 
narrow function for $P=\Lambda$ and  becomes almost constant for $P=10 \Lambda$.

\begin{widetext}

\begin{figure}[t]
    \centerline{\includegraphics[width=3in]{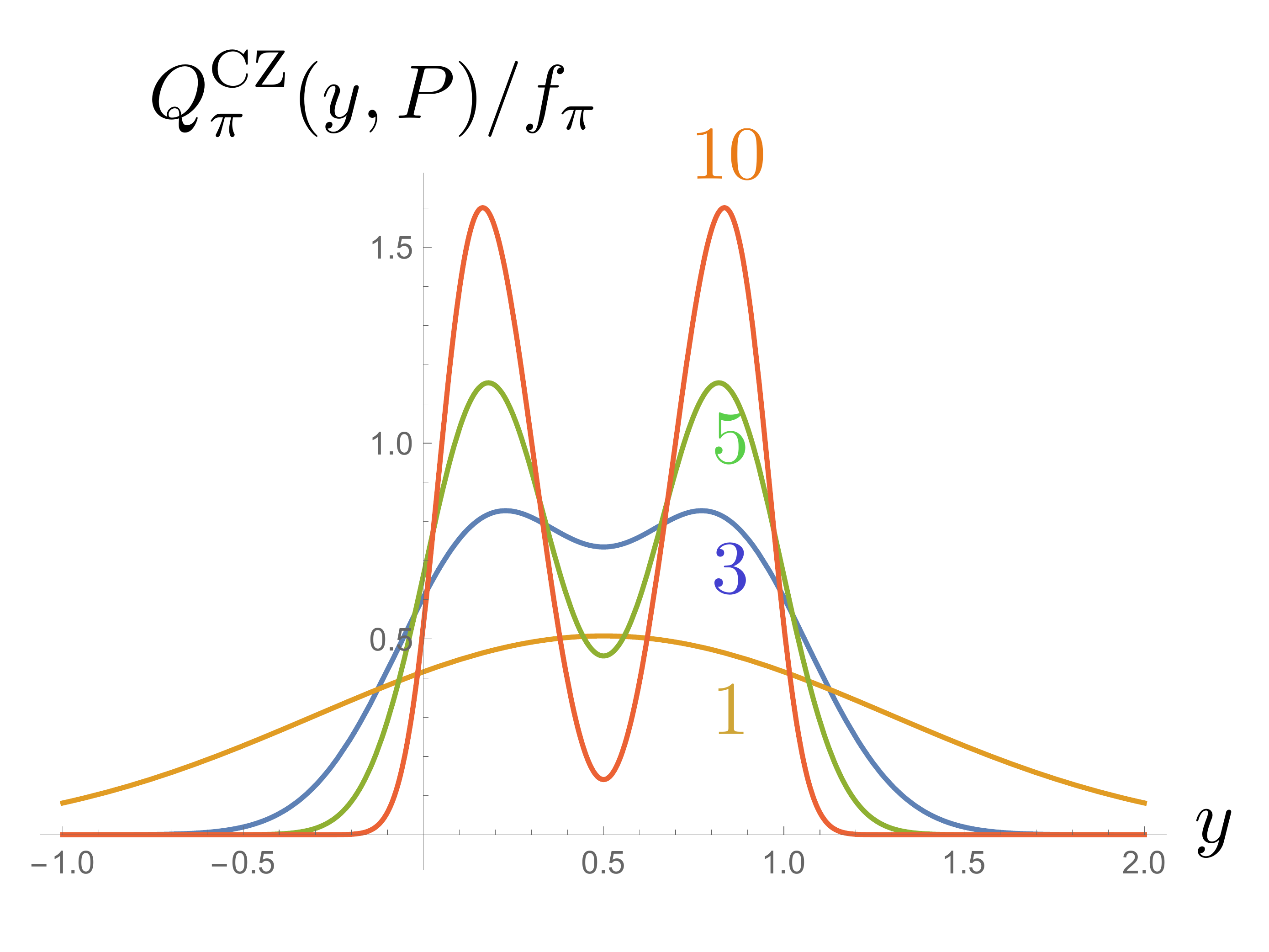} \hspace{10mm} \includegraphics[width=3in]{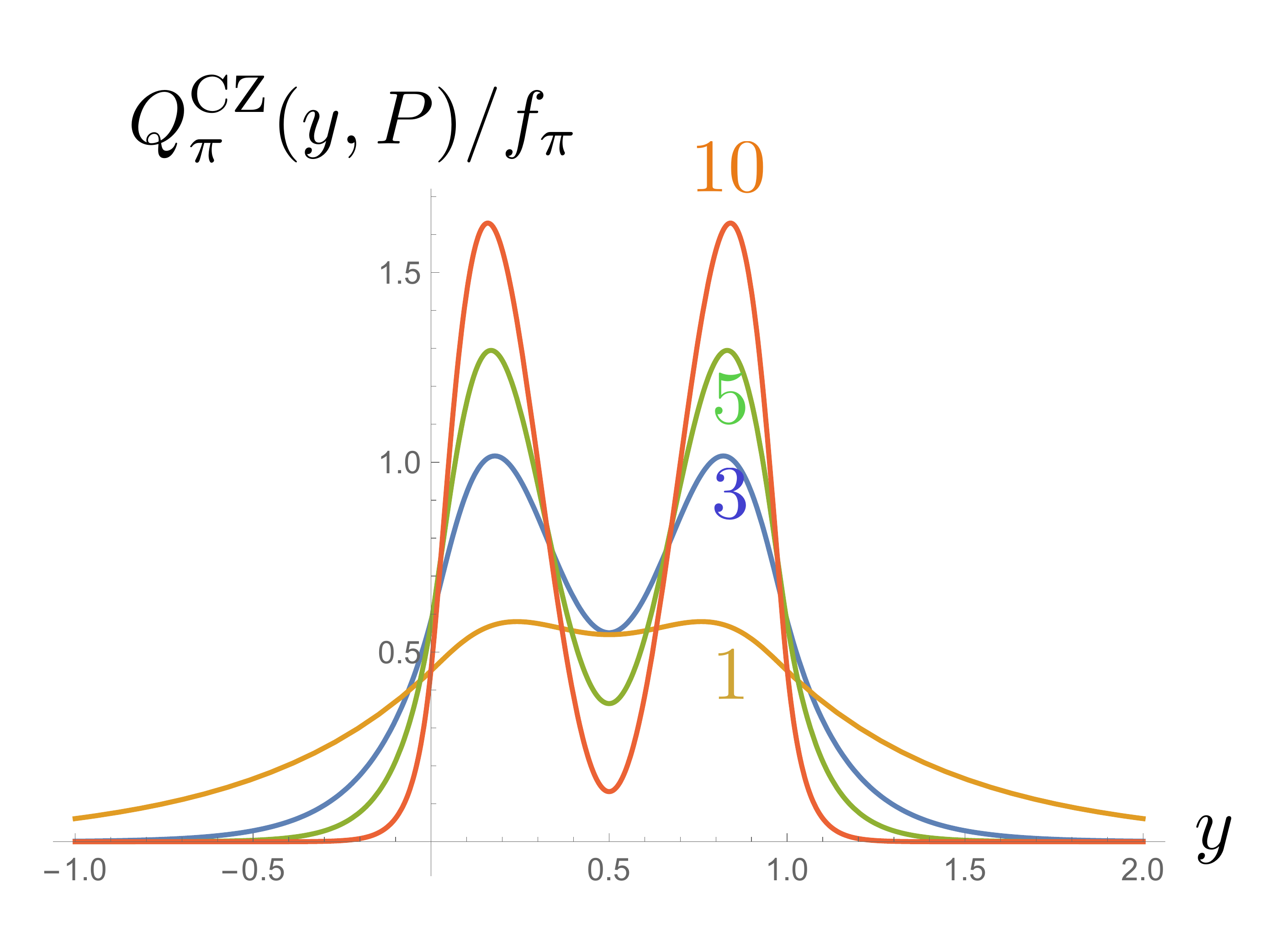}}
    \caption{Quasi-distribution amplitude   $Q^{\rm CZ}_\pi (y,P)$ for  \mbox{$P/\Lambda =1,3,5,10$}
     in the Gaussian   (left) and   ``slow''  models (right).
    \label{CZ}}
    \end{figure}

\begin{figure}[htb]
    \centerline{\includegraphics[width=3in]{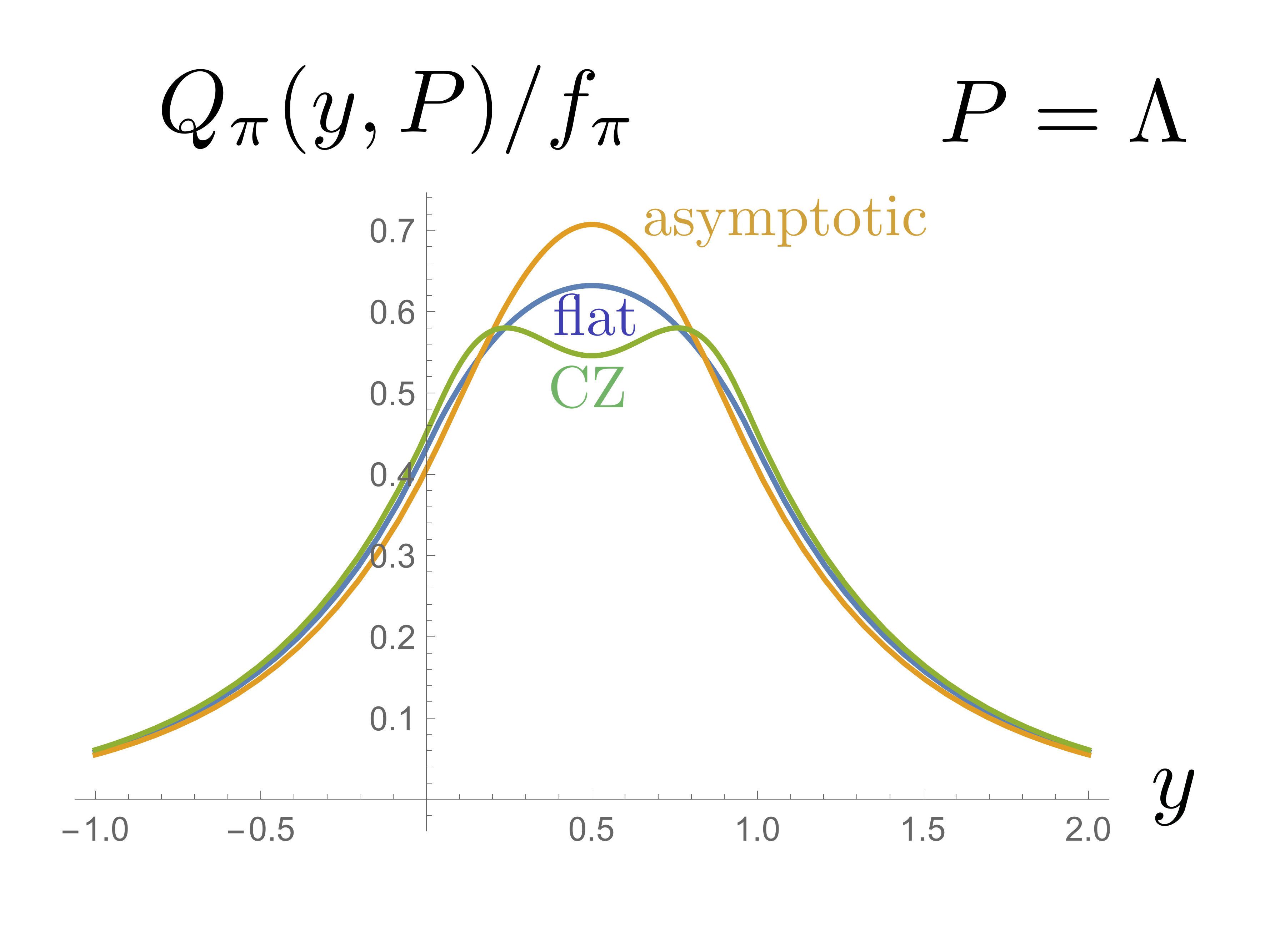}  \hspace{10mm}  \includegraphics[width=3in]{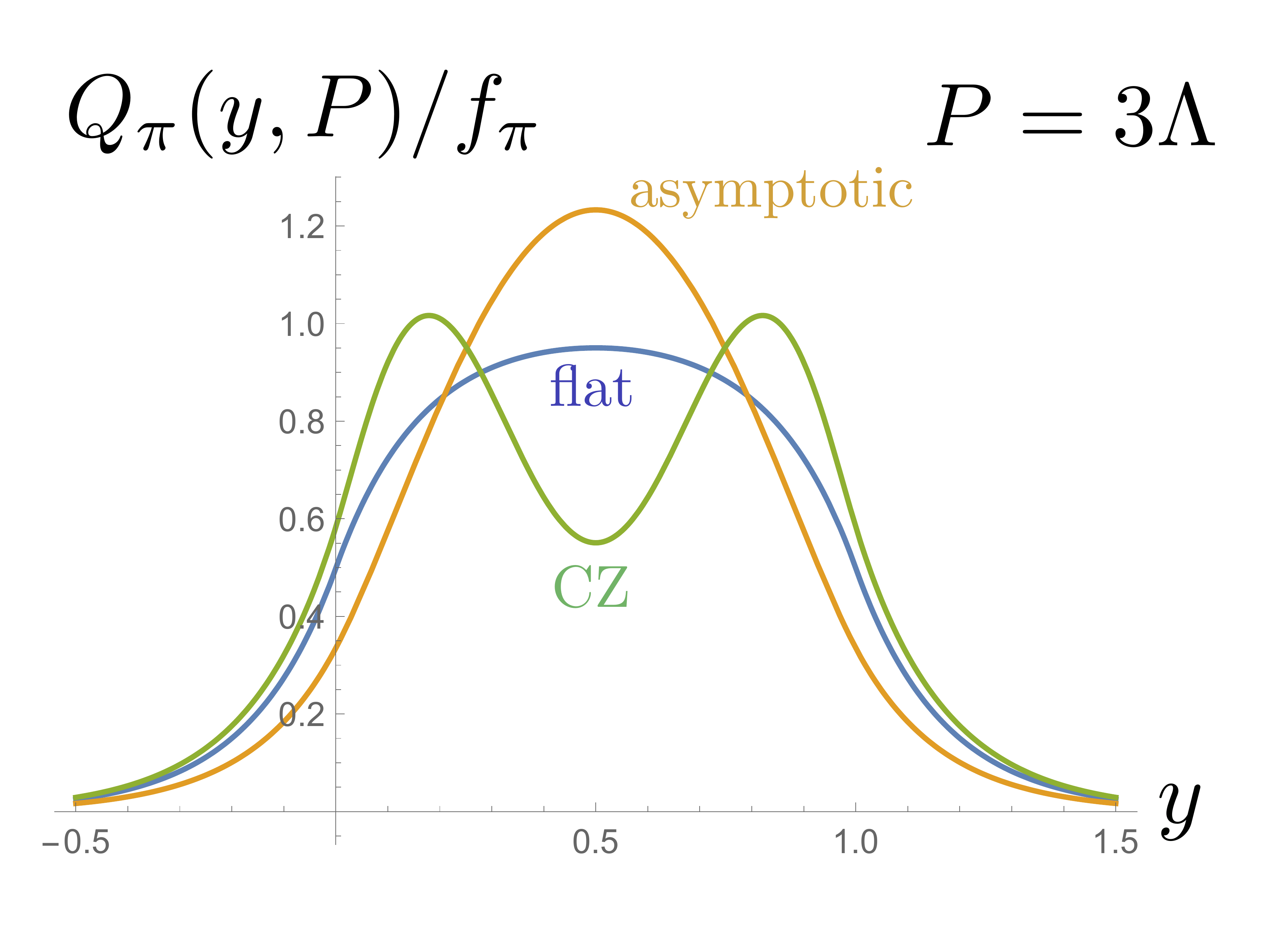}}
    \caption{Quasi-distribution amplitudes  $Q_\pi (y,P)$  in the ``slow''  model  for \mbox{$P=\Lambda $}
    (left) and  \mbox{$P= 3\Lambda $} (right)
    evolving to CZ, flat and asymptotic  DAs. 
    \label{Qfy13}}
    \end{figure}

    \begin{figure}[htb]
    \centerline{\includegraphics[width=3in]{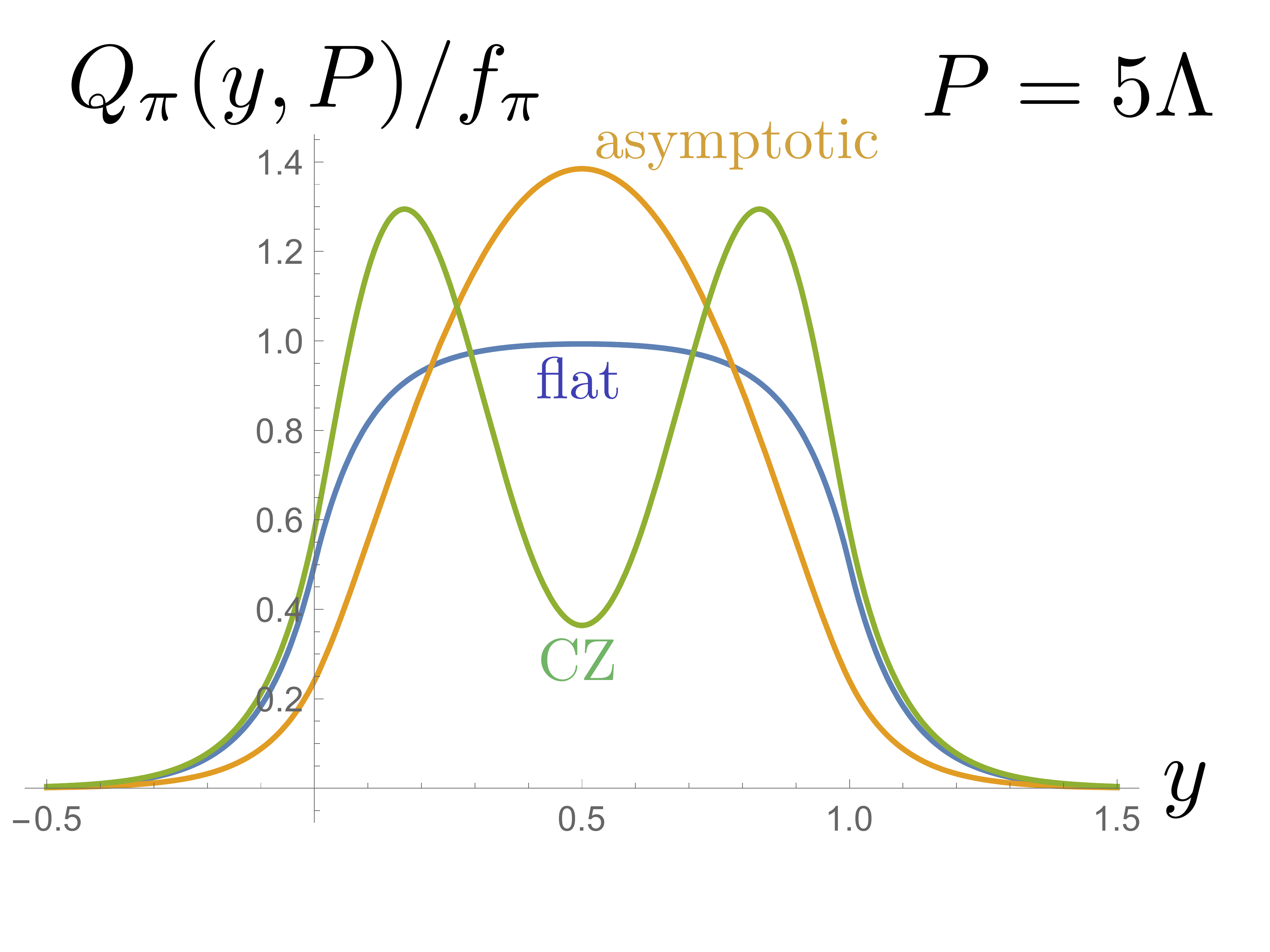} \hspace{10mm} \includegraphics[width=3in]{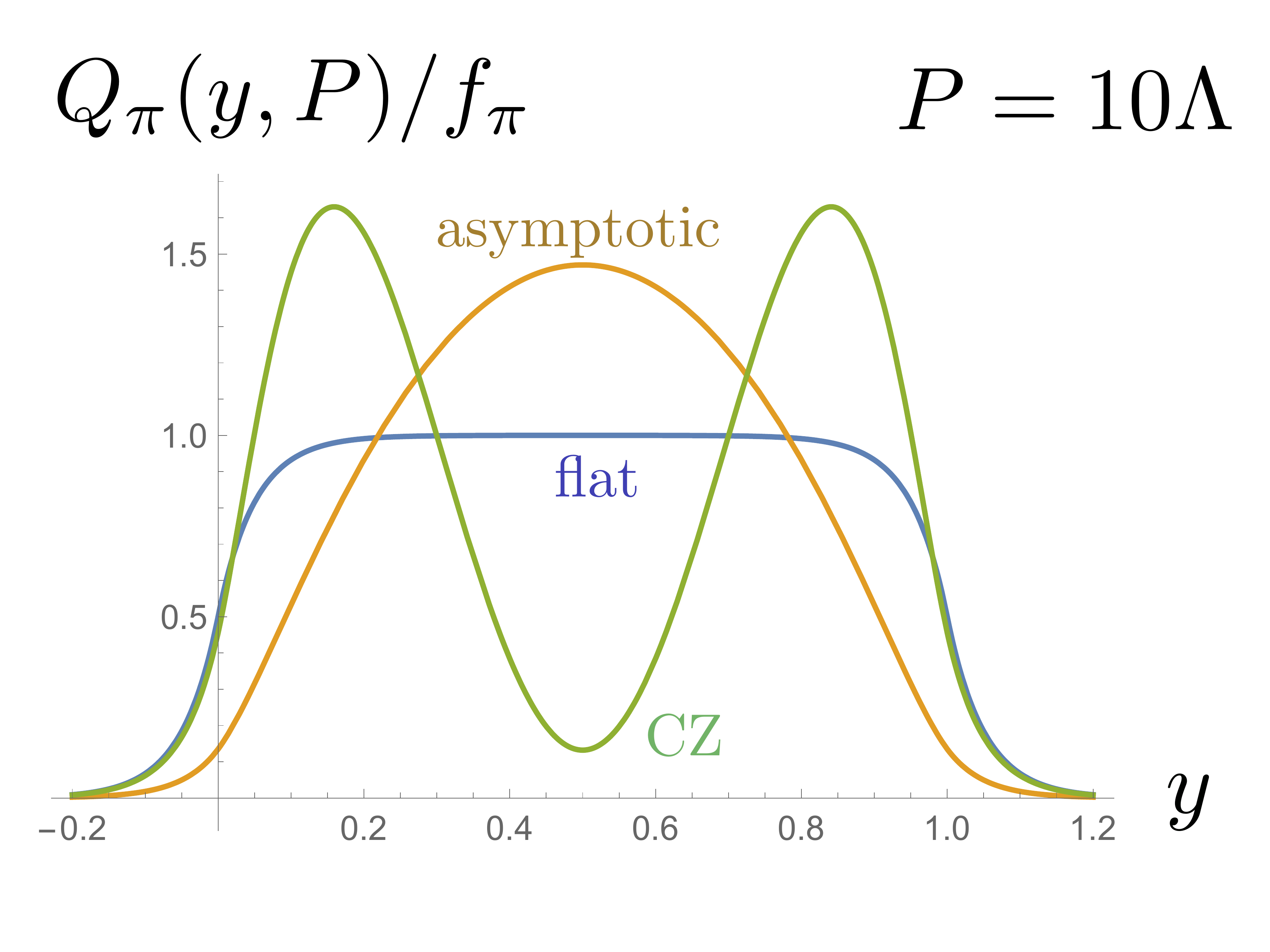}}
    \caption{Quasi-distribution amplitudes $Q_\pi (y,P)$  in the ``slow'' model  for \mbox{$P= 5 \Lambda $}
    (left) and  \mbox{$P= 10 \Lambda $} (right)
    evolving to CZ,  flat and  asymptotic DAs. 
    \label{Qfy510}}
    \end{figure}
       \end{widetext} 
    
  \section{Leading-order hard tail}

The nonperturbative evolution of $Q_\pi (y,P)$ essentially stops
for  $P/\Lambda \gtrsim 20$, and for larger values of $P$ the 
dominant role is played by  the perturbative evolution induced by the hard part. 
In our papers \cite{Radyushkin:2014vla,Radyushkin:2015gpa}, 
it was  suggested to 
take  a purely soft TMDA  (or VDA) as a starting approximation, and then ``generate'' hard tail 
by adding one-gluon exchanges. The only new parameter is the overall factor  $\alpha_s$,
while the $k_\perp$-dependence of the hard tail of the TMDA $\Psi (x,k_\perp^2)$ is 
completely determined by the soft part. 

For large $k_\perp$, the generated hard part of the TMDA has 
a $\sim 1/k_\perp^2$ behavior, but its  explicit functional form is much more complicated.
In particular, it  is finite in the $k_\perp \to 0$   limit
\cite{Radyushkin:2015gpa}.
The infrared cut-off for the naive $1/k_\perp^2$ extrapolation  is provided by the finite size of the pion
encoded in the parameters, like $\Lambda$, present in the soft TMDA.
Postponing the  analysis   of the interplay between the nonperturbative and
perturbative evolution for future studies, we just outline below
the VDA treatment of the hard tail.   

For large $\sigma$, the lowest-order (in $\alpha_s$)  hard tail has the form 
\begin{align}
   \Phi^{\rm hard}  (x,\sigma) \, = {\Delta (x)}/{\sigma} \,  
 \  , 
 \label{Phihard}
\end{align} 
 with  $\Delta (x)$  given by
 \begin{align}
 \Delta (x) =   \int_0^1dz  \, V(x,z) \, \varphi_\pi ^{\rm soft}(z) 
 \ , 
 \label{deltaev}
\end{align} 
where $V(x,z)$ is the perturbative evolution kernel   \mbox{\cite{Efremov:1978rn,Efremov:1979qk,Lepage:1979zb}.}
 The asymptotic form  (\ref{Phihard})     
corresponds to a \mbox{$\sim 1/k_\perp^2$}  TMDA, which  is singular for $k_\perp=0$.
As explained above, this singularity is absent in the exact (rather  complicated) expression for the hard tail. 
 For illustration purposes, we take now the simplest regularization
\mbox{$ 1/k_\perp^2 \to 1/(k_\perp^2 +m^2) $}. It   corresponds to the change 
$1/\sigma \to e^{-i m^2/\sigma}/\sigma$ in the hard part of VDA, 
 \begin{align}
   \Phi^{\rm hard}  (x, \sigma) \, \to \frac{\Delta(x)}{\sigma}e^{- im^2/\sigma}  \,  
 \ . 
 \label{Phihardm}
\end{align}  
To proceed with the conversion formula, one needs the integral over $\sigma$
 \begin{align}
 I  (x,y, P)  = & \,\int_{0}^{\infty} \frac{d   \sigma}{ \sqrt{\pi \sigma}}
\frac{P}{\sigma}   \,   
  e^{- (x -y)^2 P^2 / \sigma- m^2/\sigma } \nonumber \\
  &=  \frac{1}{\sqrt{(x -y)^2+m^2/ P^2}}
 \ . 
 \label{hardsigma}
\end{align} 
This gives the hard part of the quasi-distribution  amplitude 
 \begin{align}
 Q_\pi^{\rm hard}  (y, P)  = &\,
 \int_{0}^1 dx\,  %
 \frac{\Delta (x)}{   
  \sqrt{(x -y)^2 +m^2/P^2} }
 \ . 
 \label{deltaQ}
\end{align} 
It generates  evolution with respect to $P^2$ in the form 
\begin{align}
P^2 \frac{d}{dP^2}  Q^{\rm hard}_\pi    (y, P)  = & \, \frac{ m^2}{2 P^2} 
 \int_{0}^1 dx\,  %
 \frac{\Delta (x)}{   
  [{(x -y)^2 +m^2/P^2}]^{3/2} }
 \ . 
 \label{evoQ}
\end{align} 
Taking   the $m/P \to 0$ limit we have 
\begin{align}
 \frac{ m^2}{2 P^2} &
 \int_{0}^1 dx\,  %
 \frac{V (x,z)}{   
  [{(x -y)^2 +m^2/P^2}]^{3/2} }
  \nonumber \\ & = \,V (y,z)   +
 {\cal O} (m^2/P^2)
 \ , 
 \label{evoQ2}
\end{align} 
i.e. for large $P^2$ the quasi-distribution amplitude  evolves according to the 
 perturbative evolution  equation with respect to  $P^2$.

\section{Conclusions}

In this paper, we extended  the approach 
of Ref. \cite{Radyushkin:2016hsy},  where we have been dealing 
 with the parton distribution functions,
the basic ingredients of perturbative QCD analysis of hard inclusive processes.
Now we have dealt with the  pion distribution amplitude, the basic ingredient of hard exclusive 
processes involving the pion. We 
applied the formalism of  virtuality distribution amplitudes 
to study the $p_3$-dependence of quasi-distribution amplitudes $Q_\pi (y,p_3)$.

Just like in Ref.   \cite{Radyushkin:2016hsy},  we have 
established  a simple relation between   QDAs and TMDAs  that allows 
to derive models for QDAs from the models for  TMDAs.
Unlike  the PDF case, there are many drastically  
different models claimed to describe  the primordial 
shape of the pion DA.  We  have presented  the $p_3$-evolution patterns 
for models producing  some popular proposals: Chernyak-Zhitnitsky, flat and asymptotic DAs.
Our results 
 may be used as a  guide  for future  studies 
 of the pion distribution amplitude   on  the lattice using the quasi-distribution approach. 
 
As our  estimates show, one would need $P$ of the order of a few
 GeV for  the nonperturbative evolution to settle.  It is natural to expect that
 perturbative evolution will be rather important at such scales. Thus, 
an interesting and technically challenging question for future studies is the interplay 
 between the nonperturbutive and perturbative evolution of the 
 pion quasi-distribution amplitude.

\section*{Acknowledgements}

This work is supported by Jefferson Science Associates,
 LLC under  U.S. DOE Contract \#DE-AC05-06OR23177
 and by U.S. DOE Grant \#DE-FG02-97ER41028.




\begin{thebibliography}{11}

\bibitem{Braun:2015axa} 
  V.~M.~Braun, S.~Collins, M.~G{\"o}ckeler, P.~P{\'e}rez-Rubio, A.~Sch{\"a}fer, R.~W.~Schiel and A.~Sternbeck,
  Phys.\ Rev.\ D {\bf 92}, no. 1, 014504 (2015).



\bibitem{Ji:2013dva} 
  X.~Ji,
  Phys.\ Rev.\ Lett.\  {\bf 110}, 262002 (2013).



\bibitem{Lin:2014yra} 
  H.~W.~Lin,
  PoS LATTICE {\bf 2013}, 293 (2014).



\bibitem{Alexandrou:2015rja} 
  C.~Alexandrou, K.~Cichy, V.~Drach, E.~Garcia-Ramos, K.~Hadjiyiannakou, K.~Jansen, F.~Steffens and C.~Wiese,
  Phys.\ Rev.\ D {\bf 92}, 014502 (2015).



\bibitem{Lin:2015vxw} 
  H.~W.~Lin,
  Few Body Syst.\  {\bf 56}, no. 6-9, 455 (2015).



\bibitem{Chen:2016utp} 
  J.~W.~Chen, S.~D.~Cohen, X.~Ji, H.~W.~Lin and J.~H.~Zhang,
  Nucl.\ Phys.\ B {\bf 911}, 246 (2016). 



\bibitem{Alexandrou:2016bud} 
  C.~Alexandrou, K.~Cichy, K.~Hadjiyiannakou, K.~Jansen, F.~Steffens and C.~Wiese,
  PoS DIS {\bf 2016}, 042 (2016).



\bibitem{Alexandrou:2016tjj} 
  C.~Alexandrou,
  Few Body Syst.\  {\bf 57}, no. 8, 621 (2016).

\bibitem{Xiong:2013bka} 
  X.~Xiong, X.~Ji, J.~H.~Zhang and Y.~Zhao,
  Phys.\ Rev.\ D {\bf 90}, no. 1, 014051 (2014)


\bibitem{Gribov:1972ri} 
  V.~N.~Gribov and L.~N.~Lipatov,
  Sov.\ J.\ Nucl.\ Phys.\  {\bf 15}, 438 (1972)
  [Yad.\ Fiz.\  {\bf 15}, 781 (1972)].



\bibitem{Lipatov:1974qm} 
  L.~N.~Lipatov,
  Sov.\ J.\ Nucl.\ Phys.\  {\bf 20}, 94 (1975)
  [Yad.\ Fiz.\  {\bf 20}, 181 (1974)].



\bibitem{Altarelli:1977zs} 
  G.~Altarelli and G.~Parisi,
  Nucl.\ Phys.\ B {\bf 126}, 298 (1977).



\bibitem{Dokshitzer:1977sg} 
  Y.~L.~Dokshitzer,
  Sov.\ Phys.\ JETP {\bf 46}, 641 (1977)
  [Zh.\ Eksp.\ Teor.\ Fiz.\  {\bf 73}, 1216 (1977)].



\bibitem{Radyushkin:2016hsy} 
  A.~Radyushkin,
  arXiv:1612.05170 [hep-ph].



\bibitem{Radyushkin:2014vla} 
  A.~V.~Radyushkin,
  Phys.\ Lett.\ B {\bf 735}, 417 (2014). 



\bibitem{Radyushkin:2015gpa} 
  A.~V.~Radyushkin,
  Phys.\ Rev.\ D {\bf 93}, no. 5, 056002 (2016).



\bibitem{Gamberg:2014zwa} 
  L.~Gamberg, Z.~B.~Kang, I.~Vitev and H.~Xing,
  Phys.\ Lett.\ B {\bf 743}, 112 (2015).



\bibitem{Gamberg:2015opc} 
  I.~Vitev, L.~Gamberg, Z.~Kang and H.~Xing,
  PoS QCDEV {\bf 2015}, 045 (2015).



\bibitem{Bacchetta:2016zjm} 
  A.~Bacchetta, M.~Radici, B.~Pasquini and X.~Xiong,
  arXiv:1608.07638 [hep-ph].



\bibitem{Radyushkin:1977gp} 
  A.~V.~Radyushkin,
  JINR-P2-10717 (1977); English translation: 
  arXiv:hep-ph/0410276.



\bibitem{Efremov:1978rn} 
  A.~V.~Efremov and A.~V.~Radyushkin,
  Theor.\ Math.\ Phys.\  {\bf 42}, 97 (1980)
  [Teor.\ Mat.\ Fiz.\  {\bf 42}, 147 (1980)].



\bibitem{Efremov:1979qk} 
  A.~V.~Efremov and A.~V.~Radyushkin,
  Phys.\ Lett.\  {\bf 94B}, 245 (1980).



\bibitem{Lepage:1979zb} 
  G.~P.~Lepage and S.~J.~Brodsky,
  Phys.\ Lett.\  {\bf 87B}, 359 (1979).


\bibitem{Efremov:1978fi}   A.~V.~Efremov and A.~V.~Radyushkin,
{\it High Momentum Transfer Processes in QCD}.
 JINR-E2-11535, Apr 1978. 30pp.
Submitted to 19th Int. Conf. on High Energy Physics, Tokyo, Japan, Aug 23-30, 1978. 

\bibitem{Chernyak:1981zz} 
  V.~L.~Chernyak and A.~R.~Zhitnitsky,
  Nucl.\ Phys.\ B {\bf 201}, 492 (1982)
  Erratum: [Nucl.\ Phys.\ B {\bf 214}, 547 (1983)].



\bibitem{Dittes:1981aw} 
  F.~M.~Dittes and A.~V.~Radyushkin,
  Sov.\ J.\ Nucl.\ Phys.\  {\bf 34}, 293 (1981)
  [Yad.\ Fiz.\  {\bf 34}, 529 (1981)].

\bibitem{Anikin:1999cx} 
  I.~V.~Anikin, A.~E.~Dorokhov and L.~Tomio,
  Phys.\ Lett.\ B {\bf 475}, 361 (2000).


\bibitem{RuizArriola:2002bp} 
  E.~Ruiz Arriola and W.~Broniowski,
  Phys.\ Rev.\ D {\bf 66}, 094016 (2002).



\bibitem{Radyushkin:2009zg} 
  A.~V.~Radyushkin,
  Phys.\ Rev.\ D {\bf 80}, 094009 (2009).



\bibitem{Polyakov:2009je} 
  M.~V.~Polyakov,
  JETP Lett.\  {\bf 90}, 228 (2009).



\bibitem{Mikhailov:1986be} 
  S.~V.~Mikhailov and A.~V.~Radyushkin,
  JETP Lett.\  {\bf 43}, 712 (1986)
  [Pisma Zh.\ Eksp.\ Teor.\ Fiz.\  {\bf 43}, 551 (1986)].

\bibitem{Lepage:1980fj} 
  G.~P.~Lepage and S.~J.~Brodsky,
  Phys.\ Rev.\ D {\bf 22}, 2157 (1980).


\bibitem{Musatov:1997pu} 
  I.~V.~Musatov and A.~V.~Radyushkin,
  Phys.\ Rev.\ D {\bf 56}, 2713 (1997).





\end{thebibliography}
\end{document}